# FreeBeacon: Efficient Communication and Data Aggregation in Battery-Free IoT

Gaosheng Liu, Kasım Sinan Yıldırım, Lin Wang

*Abstract*—To improve sustainability, Internet-of-Things (IoT) is increasingly adopting battery-free devices powered by ambient energy scavenged from the environment. The unpredictable availability of ambient energy leads to device intermittency, bringing critical challenges to device communication and related fundamental operations like data aggregation.

We propose FreeBeacon, a novel scheme for efficient communication and data aggregation in battery-free IoT. We argue that the communication challenge between battery-free devices originates from the complete uncertainty of the environment. FreeBeacon is built on the insight that by introducing just a small degree of certainty into the system, the communication problem can be largely simplified. To this end, FreeBeacon first introduces a small number of battery-powered devices as beacons for battery-free devices. Then, FreeBeacon features protocols for battery-free devices to achieve interaction with the beacon and to perform communication efficiently following customized schedules that implement different data aggregation schemes while achieving resilience. We evaluate FreeBeacon with extensive prototype-based experiments and simulation stud- ies. Results show that FreeBeacon can consistently achieve an order of magnitude data aggregation efficiency when compared with the state-of-the-art approaches.

## I. INTRODUCTION

Internet-of-Things (IoT) has been widely deployed to sup- port various smart applications like environment monitoring, precision agriculture, and healthcare [1], [2], [3], [4]. Yet, traditional IoT systems built with battery-powered devices require significant maintenance efforts and bring environmen- tal pollution threats, all attributed to the onboard chemical battery [5], [6]. To address this issue, researchers have ex- plored the possibility of building IoT systems with battery- free devices that scavenge ambient energy to sustain their operations [7], [8], [9], [10]. By removing the battery from IoT devices, modern IoT systems can be made self-sustainable.

However, the uncertain ambient energy availability brings significant challenges to battery-free IoT. Due to the insuf- ficient and likely also unstable energy supply, battery-free devices typically employ a small capacitor to buffer the scav- enged energy and work intermittently—waking up to execute the program (for a few milliseconds) when the capacitor voltage reaches a high threshold and falling asleep to recharge the capacitor (for 10s–100s of ms) when the voltage drops below a low threshold [11], [12], as depicted in Figure 1. To handle such intermittency, existing works have focused on task- and checkpointing-based mechanisms to ensure the forward progress of program execution [13], [14], [15], [16], [17], [18], [19].

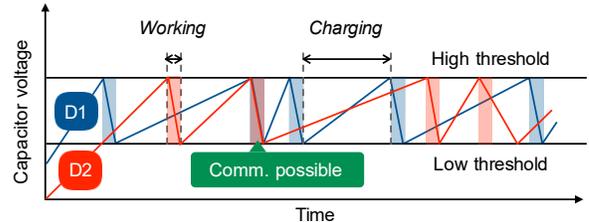

Fig. 1: Intermittency of battery-free devices with varying charging time. Devices D1 and D2 can communicate only when they both work simultaneously.

Intermittency also presents significant challenges for wire- less communication among battery-free IoT devices, rendering it unreliable with very low and highly unpredictable success rates [11], [10], [12], [20]. Specifically, charging time is uncontrollable, making essential network operations such as routing and data aggregation very difficult. In particular, we identify the following core challenges:

- **Device discovery (C1):** Battery-free devices can have long yet varying and unpredictable charging times, often inter- leaving their short active periods. Therefore, it is hard for two devices to wake up simultaneously, discover each other, and exchange data. Even if two devices encounter, they quickly lose synchrony, making pairwise communication highly unsuccessful.

- **Structured communication (C2):** The challenge of device discovery complicates essential network operations signifi- cantly. For instance, tasks like data aggregation often require specific device pairs to follow specific communication pat- terns (e.g., a tree structure), which relies on reliable pairwise communication. Furthermore, frequent device failures make it difficult to maintain such communication patterns since when a device failure happens, battery-free devices lose their hardware/software states (e.g., lose their notion of time). Therefore, it is essential to recover and restore the communication patterns gracefully.

Existing works mainly focus on the device discovery prob- lem (C1), but a generalizable solution is still missing. Besides, establishing structured communication (C2) in battery-free IoT has yet to be explored. For instance, to address the interleaving problem in device discovery, Find [11] introduces random delays to device wake-up times using statistical charging time distributions with fine-tuned parameters to minimize discovery latency. However, tuned parameters may quickly become inef- fective and lead to poor performance when the environmental conditions and charging time change. Some other studies leverage external shared signals (e.g., powerline flicker [11]) or low-power RF sensing [21] to expedite synchronization.

However, they all require special hardware, and some only work in restricted environments (e.g., where shared indoor lighting over all devices is possible). Importantly, none of these existing works address the challenges C1 and C2 jointly. As a result, efficient and structured communication in battery-free IoT remains an unsolved critical problem.

This paper presents FreeBeacon, a novel scheme for efficient networking in battery-free IoT. Considering that charging time uncertainty is the main culprit to inefficiency, our key insight is to simplify battery-free communication by introducing a *small degree of certainty* artificially. Based on this insight, we propose to build a hybrid IoT system where we place a small number of battery-powered devices among battery-free devices. Those battery-powered devices serve as *beacons* sharing critical information and guiding the communication between battery-free devices. The small number of battery-powered devices can be easily deployed and maintained, with minimum impact on sustainability.

FreeBeacon exploits the certainty provided by beacons to facilitate communication with a two-phase design. In the first phase, FreeBeacon features a beacon discovery protocol for battery-free devices to contact a beacon resiliently. To this end, FreeBeacon instructs the beacon and all battery-free devices to follow fixed-length cycles (through delaying), respectively, where the length of the two cycles are co-prime. Following the Weyl sequence theory [22], FreeBeacon ensures that each battery-free device makes persistent contact with its assigned beacon periodically (C1). The beacon broadcasts critical global information that helps each battery-free device stay in its pre-allocated slots to simplify communication. In the second phase, devices communicate directly based on the globally shared slot allocation and follow scheduled communication orders (C2) to implement data aggregation schemes. FreeBeacon is resilient to device failures (where the device is completely reset) by construction. When a battery-free device recovers from a device failure, it is guaranteed to meet the beacon as long as it follows the pre-defined cycle. Upon contact with the beacon, the battery-free device can receive the communication schedule from the beacon and restore the communication pattern.

Overall, we make the following contributions:

- We propose, for the first time, a hybrid design for sustainable IoT where we leverage the little certainty provided by a few battery-powered devices to simplify the communication and data aggregation among battery-free devices with failure resilience.
- We present FreeBeacon, an efficient and robust scheme for beacon discovery and device communication under the hybrid design, and showcase how different data aggregation schemes can be implemented by FreeBeacon.
- We perform extensive prototype-based experiments and simulations to validate the performance of FreeBeacon under varying scenarios and setups. Results show that FreeBeacon can achieve up to 29.5× performance gains for pairwise communication and data aggregation when compared with the retrofitted state-of-the-art approaches for communication in battery-free IoT.

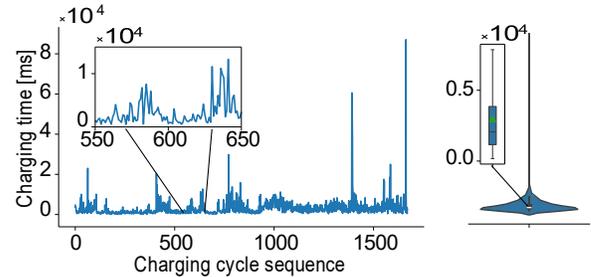

Fig. 2: High variability in the charging time of a battery-free IoT device equipped with piezoelectric harvesters on cars driving on different roads [23].

## II. BATTERY-FREE IOT NETWORKING

Battery-free devices leverage tiny capacitors to buffer scavenged energy and they operate intermittently, alternating between charging and working states across charging cycles, as shown in Figure 1. The working time in each charging cycle is mostly constant under a given hardware specification and IoT application, but the charging time is dictated by the energy availability of the ambient environment (e.g., light intensity, RF signal strength) and can be highly variable. To quantify the variability level of the charging time, we dive into the charging time trace released in a public dataset [23], which is collected from three piezoelectric harvesters mounted on the windshield, dashboard, and in the trunk of two cars with a 17 $\mu$F capacitor while the cars drive for two hours in a convoy over different roads. Figure 2 depicts (left) a snippet of the evolution of the charging time across multiple charging cycles and (right) the charging time distribution. As we can observe, the charging time is widely distributed where the maximum-to-minimum ratio can reach 509 and the ratio between the 3rd quartile and the 1st quantile can be as large as 3.3. More importantly, such variability is quite random and is hence hard to predict.

### A. Intermittent Communication

The charging time variability brings significant challenges to the device communication problem in battery-free IoT. With traditional battery-powered devices (even when they are duty-cycled), a device has a clear expectation when its communicating counterpart is active. This allows for scheduled communication to happen easily. However, this condition does not hold anymore when communication happens between two battery-free devices. Due to the short yet highly variable charging time, two devices operate intermittently and will have no shared knowledge about when they will become active simultaneously—a necessary condition for communication. More specifically, we identify two major challenges in intermittent communication.

**Challenge of device discovery.** Battery-free devices must discover each other first before they can initiate communication. Focusing on this device discovery problem, existing works like Find [11] follow the random-guess-based idea where they propose to introduce an artificial delay to extend the charging time of a battery-free device, hoping that another battery-free device happens to wake up at the same time with also an artificial delay applied. The artificial delay is generated following a statistical distribution with fine-tuned parameters so that the device discovery time is minimized in expectation. However, the challenge arises from tuning the parameters of the statistical distribution according to specific scenarios. For example, when the environmental energy condition changes, the device charging time distribution also drifts. Consequently, the tuned parameters may quickly become ineffective, leading to suboptimal performance. To verify this point, we compare the performance of Find under tuned and non-tuned parameters for a geometric distribution. For the tuned case, we pick the parameters optimized by Find for a specific scenario (charging time randomly distributed in range [501, 1000] slots), while for the non-tuned case, we use parameters for a different scenario (charging time randomly distributed in range [1, 500] slots). Figure 3a demonstrates the big gap in discovery time between the two cases, highlighting the generalizability issue of the random-guess-based approach.

**Challenge of structured communication.** IoT communication typically goes way beyond discovering devices randomly and require to follow structured communication patterns where device pairs must communicate in a pre-specified manner. For example, data aggregation requires a set of IoT devices to form a specific topology (e.g., a line or a tree) and communication must happen along the edges of the topology, often also meeting specific timing conditions. This exacerbates the challenge posed by the aforementioned charging time variability since now a device has to discover the exact device to communicate to, not just a random device. One can build a structured communication protocol based on unstructured discovery, but the communication efficiency would soon become impractical as the involved devices grow. To verify this point, we perform a small experiment where we compare the number of successful communications among varying numbers of battery-free devices which follow the charging time traces depicted in Figure 2 within a time limit of 300K seconds, under two different conditions: (1) Random where devices randomly form pairs and (2) Scheduled where devices must form pairs according to a pre-defined pair schedule. Figure 3b shows that for random pairing the success rate increases with the number of devices due to a larger probability of discovery, while it drops for the Scheduled case, highlighting the efficiency issue of the random-guess-based approach in structured communication. Additionally, battery-free devices frequently suffer from device failures with complete device reset. Being resilient to such failures—being able to quickly restore the communication pattern after device recovery—is of vital importance.

### B. The Power of (Little) Certainty

The above analysis inspires us to rethink the tradeoff point between system sustainability and efficiency in a future IoT system. More specifically, we ask the following question: *"Is it really worth it to build an IoT system completely from battery-free devices to fulfill the sustainability promise, given the poor efficiency caused by environmental uncertainty?"*

We argue that slightly relaxing the fully battery-free assumption could provide us with better designs. Our insight is that by introducing a limited degree of *certainty* into the IoT system, we can achieve efficiency while maintaining (quasi-strong) sustainability. More specifically, we propose to build hybrid IoT systems where, among a large number of battery-free devices, we introduce a small set of battery-powered devices as the source of certainty. These battery-powered devices serve as *beacons* to the battery-free devices, guiding them to discover and communicate with each other in a principled manner. With such a hybrid design, (structured) device communication among battery-free devices could be greatly simplified, with significant efficiency improvement.

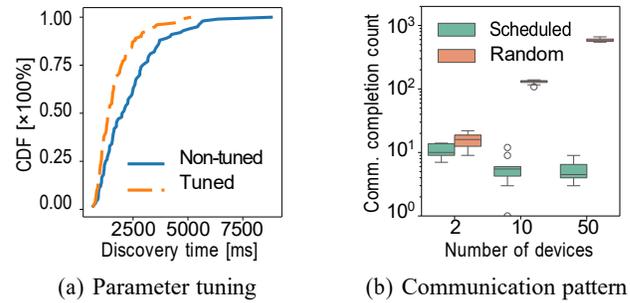

(a) Parameter tuning  (b) Communication pattern

Fig. 3: Find [11] with/without parameter tuning and under different communication patterns.

**Eliminating random guesses.** Figure 4 illustrates the high-level idea of the hybrid design with the example of one battery-powered device serving as the beacon for three battery-free devices namely D1, D2, and D3. The beacon is powered on constantly, with fixed duty-cycling known to all participating devices. In the bootstrap phase, each battery-free device first discovers the beacon [24], [20], [25]. Upon discovery, the beacon allocates a unique slot on a pre-defined cycle with a fixed number of slots to the battery-free device. The length of this cycle is shared among all battery-free devices and the slot allocation is also made known to all devices. For a battery-free device to communicate with another one, the device can jump to the allocated slot of the target on this shared cycle and successful communication is guaranteed over time. This principled approach eliminates the need for random guesses completely.

**Simplifying structured communication.** Consider a data aggregate operation with structured communication D1→D2→D3. Instead of relying on randomly guessing between D1-D2 and D2-D3 device pairs and figuring out the correct order, the beacon largely simplifies this process: D1 initializes a communication to D2 by jumping to the allocated slot of D2 directly. Within a reasonable amount of time, D1 will be able to establish communication with D2. Similarly, D2 establishes communication with D3 by jumping to the allocated slot of D3. As we can see, structured communication now follows a standard procedure and is largely simplified.

**Negligible overhead.** One may argue that the introduction of battery-powered devices into the system defeats the sustainability premise of going battery-free. We argue that this is not necessarily the case. Compared with battery-free devices, battery-powered devices are negligible in number, so the needed maintenance efforts are not in the same order of magnitude. Moreover, the location for those battery-powered devices could be chosen deliberately to ease maintenance. We believe that this hybrid design sits at a favorable tradeoff point for the design, considering the enormous gains vs. the negligible overheads.

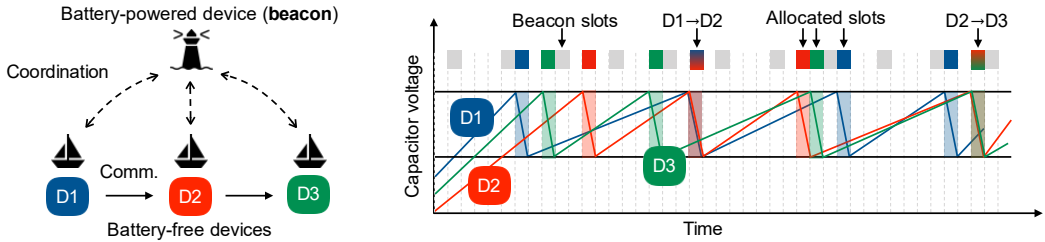

Fig. 4: Example showing how a battery-powered device serves as a beacon to facilitate structured communication (D1→D2→D3) among battery-free devices. The beacon instructs battery-free devices to stay on their allocated slots upon discovery and then a battery-free device can jump to the slot of another device for efficient communication.

## III. FreeBeacon System Design

This section presents the design of FreeBeacon, our proposed scheme for efficient communication and data aggregation in battery-free IoT following the hybrid design. FreeBeacon consists of two major components: (1) a *beacon discovery* protocol that allows battery-free devices to discover the beacon with efficiency and robustness, and (2) a *device-to-device communication* protocol that allows a battery-free device to find another device and sustain communication efficiently. Based on these protocols, FreeBeacon enables the implementation of complex structured communication patterns with scheduled device-to-device communication.

We consider the following setup for simplicity. We divide time into small fixed-length slots whose duration equals the working time of a battery-free device in a single charging cycle. While the working time of battery-free devices may vary, we can simply choose the minimum working time among all devices as the slot length. We also assume the time slot is long enough so devices can perform at least one round of communication—sending out a message and getting a response. Typically, the slot duration is a few milliseconds, while the round-tip communication can be completed within one millisecond with current hardware and wireless protocols [11], [10]. We assume one battery-powered device with duty-cycling as the beacon, which works in cycles with a fixed length $T_b$. We assume there are in total $N$ battery-free devices in the system, which is known to all devices. Each device $i \in [1...N]$ follows a cycle with varying lengths (due to varying charging times) denoted by $T_i$.

### A. Beacon Discovery

To bootstrap the communication, each battery-free device needs to establish contact with the beacon for coordination. Although the duty cycle of the beacon is manually configured and fixed, this discovery problem is not necessarily simple since the battery-free device does not know in which time slot the beacon is active.

A reasonable approach is to use linear probing, where the battery-free device extends its charging time intentionally by one slot in every charging cycle [24], [25]. The guarantee is that when the battery-free device probes more than $T_b$ slots linearly, there must be a slot in which both the beacon and the battery-free device are active. However, with multiple battery-free devices performing the same actions, collisions (where more than one battery-free device sends messages to the beacon) will happen, which will break the success guarantee. We can also use random-guess-based approaches like Find to accelerate the process, but they still suffer from efficiency and generalizability issues as already discussed.

In light of this challenge, we propose a novel beacon discovery protocol for FreeBeacon, aiming to provide high efficiency and resilience by construction. The high-level idea is to come up with a framework where the battery-free device and the beacon are **guaranteed** to meet each other eventually, regardless of possible collisions. While this seems mission-impossible at first sight, we show an interesting insight that this can be naturally achieved by leveraging an important result in number theory.

*1) Example Scenario.:* We focus on one device and consider the example shown in Figure 5. The cycle followed by the beacon has a length $T_b = 2$, i.e., the beacon wakes up every other slot. We define a *distribution cycle* $T_{dist}$ where $T_{dist} \geq N$. Each battery-free device is instructed to follow the distribution cycle, e.g., they only wake up at slots that are multiples of $T_{dist}$, achieved by deliberately extending their charging time for alignment if needed. In the example, we set $T_{dist} = 5$ so the device wakes up every five slots, as highlighted by the blue boxes.

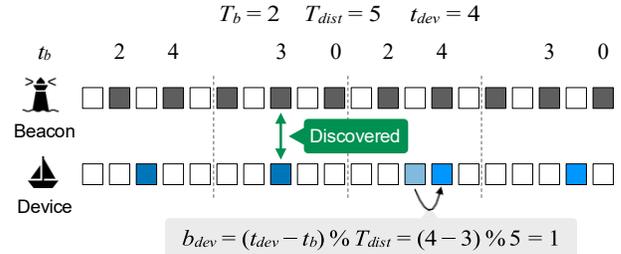

Fig. 5: Example for beacon discovery.

**Slot allocation.** On this distribution cycle, each device $i$ is allocated a unique slot $t_i \in [1...T_{dist}]$, which can be done based on the device ID uniquely allocated before deployment. Consequently, the slot allocation is shared knowledge among all battery-free devices. In this example, we assign the unique slot of the device to 4 without loss of generality, that is, $t_{dev} = 4$, where *dev* represents our specific device. Note that initially, the distribution cycle on different devices is not synchronized. As a result, even though different devices are allocated different slots on this distribution cycle, they may still wake up at the same time, causing collisions.

**Beacon index.** The beacon plays a critical role in synchronizing the distribution cycle on different devices. To this end, the beacon assumes a reference distribution cycle and keeps track of an index $t_b$ calculated as:

$$t_b = (n \times T_b)\% T_{dist} \tag{1}$$

where $n \in \mathbb{N}^+$ denotes the cycle count of the beacon after deployment. This beacon index marks the slots on the reference distribution cycle in which the beacon is active (highlighted by the shaded boxes in the figure).

**Device-beacon contact.** Each battery-free device, when waking up, broadcasts discovery messages to the beacon. When a device-beacon contact is successful, the beacon replies with the current index $t_b$. In the example, the beacon replies with $t_b = 3$ when meeting the device $dev$.

**Device synchronization.** When receiving $t_b$, battery-free device $i$ compares it with its allocated slot $t_i$. If $t_i = t_b$, the device knows that it wakes up in the correct slot on the distribution cycle; otherwise, it calculates an offset:

$$o_i = (t_i - t_b)\% T_{dist} \tag{2}$$

which represents the number of slots the device needs to wait further before it reaches its allocated slot on the distribution cycle. In the example, the device notice that $t_{dev} \neq t_b$ since $t_{dev} = 4$ while $t_b = 3$. Hence, the device calculates its offset $o_{dev} = (t_{dev} - t_b)\% T_{dist} = (4 - 3)\%5 = 1$. The device will then apply this offset to its charging time, i.e., extending its charging time by one slot in the example. This way, the device is synchronized with the beacon with respect to the distribution cycle. If all devices perform the same procedure as above, it is guaranteed that all devices will be synchronized to the same reference distribution cycle and work in unique time slots on this cycle.

*2) Guaranteed Discovery.:* In the above process, we assume that the device can make at least one successful contact with the beacon, which is non-trivial. Thankfully, the following theoretical result makes it possible.

**Theorem 1.** *If $T_b$ and $T_{dist}$ are co-prime, it is guaranteed that every device will eventually meet the beacon.*

*Proof.* See Appendix. □

Based on the above results, if all devices happen to wake up in different slots from the beginning, all devices are **guaranteed** to be discovered within $T_{dist}$ rounds. However, this condition may not hold. There might exist two or more devices that wake up in the same slot before discovering the beacon, causing collisions. To handle such cases, we introduce a simple backoff strategy where for collided devices we apply an extra delay of $T_{dist}$ slots randomly. When all battery-free devices finish the above process, it is guaranteed that all devices will only wake up in their allocated slot on the synchronized distribution cycle.

Protocol details are provided in Alg. 1 in the Appendix.

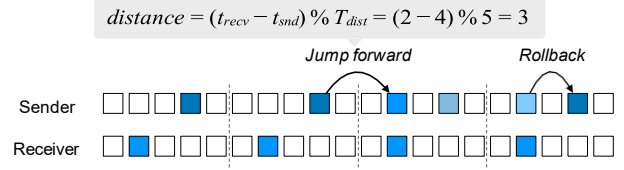

Fig. 6: Example for device-to-device communication.

### B. Device-to-Device Communication

Once all devices have discovered the beacon, communication can happen between battery-free devices in a structured manner. Specifically, every battery-free device shares the same knowledge of the allocation of slots on the distribution cycle to devices, and every device, after synchronizing the distribution cycle with the beacon, will follow this allocation to wake up periodically. Assume the system needs to perform communication between a pair of battery-free devices according to a pre-defined schedule and the two devices know their respective roles namely the *sender* and the *receiver*. The high-level idea is that the sender jumps to the receiver's slot on the distribution cycle for communication and rolls back to its own slot when the communication is done.

*1) Example Scenario.:* Consider the example shown in Figure 6, still with $T_b = 2$ and $T_{dist} = 5$. Assume two devices are serving as the sender and receiver, with allocated slots $t_{snd} = 4$ and $t_{recv} = 2$, respectively.

**Initiating communication.** Since slot allocation is shared knowledge by all battery-free devices, the sender can initiate the communication directly by jumping to the allocated slot of the receiver. To this end, the sender calculates the distance between its own slot and that of the receiver on the distribution cycle, e.g., $distance = (t_{recv} - t_{snd})\% T_{dist} = (2 - 4)\%5 = 3$, and jumps forward by extending its charging time with *distance* slots. This way, the working slots of the sender and receiver are aligned and communication can start.

**Completion.** Once the communication is completed, the sender rolls back to its originally allocated slot by reversely applying the same distance. The receiver logic is rather simple: it waits on its allocated slot and sends back an acknowledgment when receiving a valid data message.

Protocol details are provided in Alg. 2 in the Appendix.

### C. Data Aggregation

The above device-to-device communication can only be performed when the roles of the participating devices are explicitly specified. For practical IoT systems, this is typically done through a pre-defined communication pattern that specifies the sequence of communication between devices and the respective roles. Devices then periodically follow this pattern. As already discussed, data aggregation is a fundamental functionality of virtually all IoT systems. In the rest, we will focus on data aggregation and explain how three different patterns as shown in Figure 7, namely *line*, *tree*, and *ring*, can be implemented with `FreeBeacon`.

**Line-based aggregation.** We first sort all battery-free devices in an ascending order based on their IDs. As depicted in Figure 7 (left), the aggregation is initiated by the device with the smallest ID, passing its collected data to its direct neighbor with a higher ID, and so on. Note that, when a device finishes a round of communication with its neighbor, it can immediately start the next round, waiting for its neighbor to come back to the receiver role from the sender role. Therefore, a pipeline can be formed.

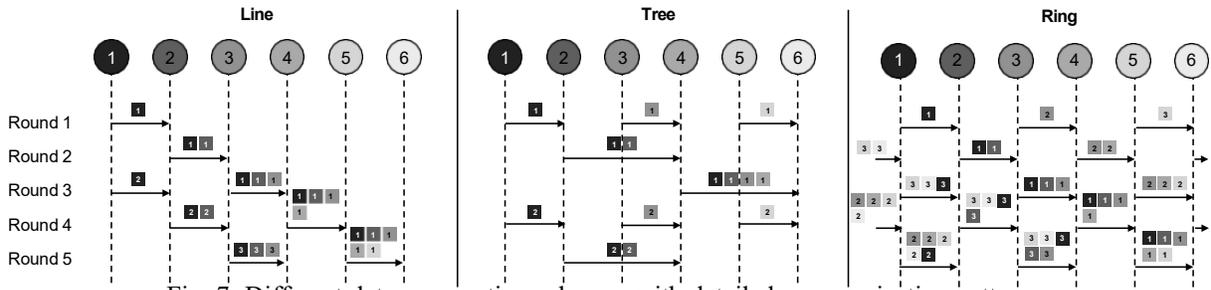

Fig. 7: Different data aggregation schemes with detailed communication patterns.

**Tree-based aggregation.** As depicted in Figure 7 (middle), we follow the simple power-of-two approach, where we always aggregate data towards devices with IDs that are powers of two (if possible) in every phase of aggregation. In the given example, the first round aggregates data to devices that can be divided by 2, the second round by 4, and so on; the last device will be used if there are not enough devices. We notice that this pattern may create a collision situation that needs to be handled. Taking device 4 as an example, devices 3 and 2 will send data to device 4 in rounds 1 and 2, respectively. However, if device 2 advances the round faster than device 3, both 2 and 3 will send data to 4 at the same time, leading to a collision. To address this problem, we propose to reverse the behavior of the sender and receiver. More specifically, we ask the receiver to jump to the slot of the sender to wait for the message. This way, for any device serving the receiver role, only the expected sender can send messages to it. In the case of our example, that means device 4 would not be able to receive the message from 2 before the message from 3 has been received and the round has been advanced.

**Ring-based aggregation.** This pattern follows the popular ring-reduce structure in collective communication where devices form a ring to pass data around and perform aggregation simultaneously on multiple data elements. The pattern for aggregating three data elements among six devices is depicted in Figure 7 (right).

In these patterns, each device only serves one role (either sender or receiver) in every round, and in the same round, different devices work in different slots on the distribution cycle. As a result, collisions are avoided by construction.

### D. Failure Resilience

A *device failure* occurs when the onboard capacitor of a battery-free device is completely drained and cannot even power the onboard real-time clock (RTC). In this case, the device completely resets and loses its sense of time. Since energy harvesting dynamics are uncontrollable, battery-free devices might fail frequently. In case of a failure, devices must repeat the discovery procedure and restore the connection to the beacon. With a high device failure rate, beacon discovery, device-to-device communication, and aggregation can easily become impractical. By construction, our protocols are **resilient** to device failures.

*1) Beacon Discovery.:* If a device experiences a device failure before a successful discovery, there is no impact on the discovery protocol. If a successful discovery has already been achieved when the device fails, the device will automatically rerun the beacon discovery protocol.

*2) Device-to-device Communication.:* During the device-to-device communication phase, the beacon helps the sender correct its working slot when a device failure occurs. When the beacon receives a message (by sniffing the broadcast channel) where the slot of the intended receiver $t_{recv}$ deviates from the current beacon index $t_b$, the beacon signals the current beacon index back asking the device to correct its slot before further communication. If the device observes the beacon being the source of the response, it means that the sender is working on the wrong slot, probably due to a recent device failure, and hence, a correction is performed by recalculating the offset with the beacon index newly announced by the beacon. When the receiver suffers from a device failure and is stuck in an incorrect slot, the beacon is not able to help correct it due to its passive role. To handle this situation, the receiver periodically sends out query messages to the beacon and applies a slot correction if needed.

*3) Data Aggregation.:* For data aggregation, there might be collisions when device failures happen. In this case, the device would stop at the expected communication step, waiting for the collision to be addressed by the slot correction mechanism in FreeBeacon. Meanwhile, the communication is paused and correctness is guaranteed. Communication pattern information and related metadata (e.g., the current sender-receiver pair) can be stored in non-volatile memory to speed up the recovery of the communication pattern.

## IV. EVALUATION

In this section, we perform comprehensive experiments to evaluate the performance of FreeBeacon under varying environmental conditions, parameter setups, and communication patterns. To this end, we implement FreeBeacon in a variety of platforms: a small-scale real-world implementation with five Riotee devices [26], a controlled testbed with microcontrollers, a numerical simulator in Python, and an implementation in the OMNeT++ simulator. Overall, we aim to answer the following questions with our evaluation.

- How does FreeBeacon perform in data aggregation?
- How efficient is the proposed beacon discovery protocol?
- How does FreeBeacon perform in other comm. tasks?

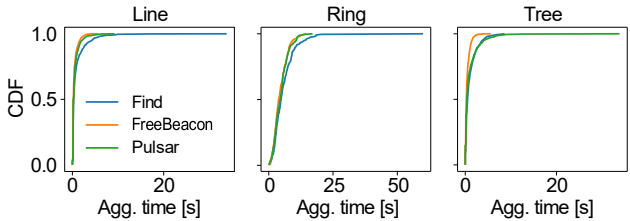

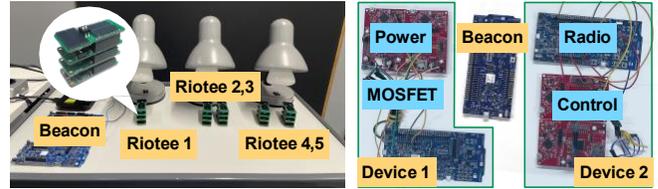

Fig. 9: (L) Riotee-base and (R) controlled testbeds.

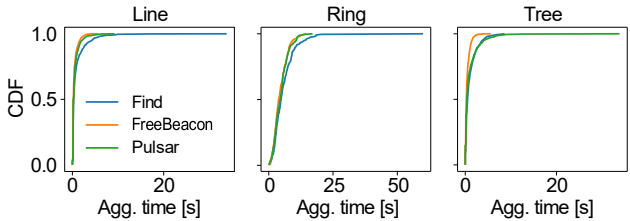

Fig. 8: Data aggregation performance of FreeBeacon.

- What are the impacts of FreeBeacon parameters?

**Common setup.** For the charging time of battery-free devices, we use both synthetic (static and random) and real-world traces (cars, jogging, office, stairs, and washer) from the public dataset [23]. In the static case, we use a fixed charging cycle of 100 slots. In the random case, the charging time is drawn from $[1, 500]$ slots uniformly at random. These numbers fall into the typical ranges reported in the traces and hence are representative. Note that the duration of the slot is given as the minimum working time of battery-free devices and hence depends on the scenario.

**Baselines.** We compare FreeBeacon against the following baselines: Find [11] where devices apply charging time delays following a fine-tuned probably distribution and look for other devices randomly, Flync-Find [11] which enhances Find by leveraging a common artificial light source and extra hardware to facilitate charging cycle alignment among battery-free devices, and Pulsar [25] which uses a coordinator to facilitate communication like FreeBeacon, but uses the linear probing approach for coordinator discovery.

### A. Overall Data Aggregation Performance

To verify the data aggregation performance of FreeBeacon under different aggregation patterns, we performance experiments based on both a small-scale real-world implementation and large-scale simulations in OMNeT++.

*1) Experiments on a Real-World Testbed.:* We first build a FreeBeacon testbed with five Riotee devices [26]. Riotee is a research platform for battery-free IoT applications, equipped with the necessary components like a solar-based energy harvester, a microcontroller, several sensors and a radio for communication. We set up the system as shown in Figure 9(left) where we use Riotee platforms as battery-free devices and one constantly powered NORDIC-NRF52840 board as the beacon. The Riotee platforms are powered by solar panels which collect energy from three Philips Hue E27 lamps (1600 lm) and store the energy in three $47\mu F$ and three $220\mu F$ capacitors. Using Seal Logic Analyzer Pro16, we measure the working time of all Riotee devices and observe that all devices have working times larger than 10 ms. Hence, we set the working time as 10 ms, which also serves as the default slot length. Based on the setup, we also measure the charging time, which mostly falls into the range of $[1, 25]$ slots.

Figure 8 depicts the average time spend for the successful aggregation of a data element across all five devices. This average is calculated as the total duration for aggregation (i.e., 10K seconds) divided by the number of successfully aggregated data elements with each of the three patterns namely line, right, and tree. We compare FreeBeacon with two baselines namely Find and Pulsar. We can notice that FreeBeacon significantly outperforms the random-guess-based approach Find with a reduction of average aggregation time at 54.73%, 23.25%, and 58.82% with the line, ring, and tree patterns, respectively. Compared with Pulsar—an idea similar to FreeBeacon but with linear probing for device discovery—FreeBeacon still achieves considerable aggregation time reduction of 23.21%, 4.8%, and 60%, respectively. The reduced gain can be explained that both FreeBeacon and Pulsar leverage a battery-powered device, confirming the effectiveness of the artificially introduced certainty. Unlike Pulsar, FreeBeacon is resilient to power failures, which we evaluate in more depth in the following sections.

*2) Simulations:* We further evaluate the performance of FreeBeacon in more diverse scenarios where we implement the three data aggregation schedules (i.e., line, tree, and ring) explained in Section III-C with the OMNeT++ simulator in version 5.6.2. We consider three ranges $[30, 120]$, $[300, 500]$, $[100, 500]$ (slots) in which we draw the charging time of battery-free devices. We assume the duration of a slot is 1 ms. We consider three different scenarios where the number of devices is set to 6, 30, and 100 respectively. The distribution cycle is set according to the number of devices and the cycle of the beacon $T_b$ is set based on the co-prime requirement. For Pulsar we use the same cycle setups for fairness. We assume each battery-free device has a total number of 1800 data items to be aggregated. We set a maximum time limit of $3 \times 10^5$ seconds for practical consideration. We consider cases with 0%, 1% and 5% device failure rates.

Table I shows the data aggregation completion time (in seconds) with different approaches and under different data aggregation patterns. For the case with 5% device failures, most of the experiments could not finish within the given time limit; hence, we omit them from the discussion.

**Line-based aggregation.** We first consider the case with no device failures. We can see that both Pulsar and Free-Beacon can complete data aggregation within the time limit under all considered scenarios. The completion time with both approaches is similar as well. In contrast, Find can only finish data aggregation in three out of nine given scenarios. Under 1% device failure rate, FreeBeacon can still finish six out of the nine scenarios, while Pulsar only manages to finish one. FreeBeacon clearly outperforms Find by a factor of 7.87 under no device failures and of 1.5 under 1% device failures. Moreover, FreeBeacon has considerable performance advantages over Pulsar in the finished scenarios. For example, in the case of charging time range $[30, 120]$ slots with six devices, FreeBeacon has improved the completion time by a factor of 6.3. Overall, FreeBeacon achieves decent

TABLE I: Data Aggregation Performance (in seconds, '–' represents incomplete) in Large-Scale Simulations

| #devices | Range | Line | | | Tree | | | Ring | | |
|---|---|---|---|---|---|---|---|---|---|---|
| | | FreeBeacon | Pulsar | Find | FreeBeacon | Pulsar | Find | FreeBeacon | Pulsar | Find |
| *No device failures* | | | | | | | | | | |
| 6 | [30, 120] | 5,230 | 5,175 | 41,203 | 6,242 | 6,076 | 36,782 | 6,597 | 52,013 | – |
| | [300, 500] | 133,052 | 131,902 | – | 153,869 | 149,926 | – | 154,322 | – | – |
| | [100, 500] | 71,849 | 73,006 | – | 84,045 | 82,692 | – | 87,863 | – | – |
| 30 | [30, 120] | 1,884 | 1,830 | 112,164 | 3,990 | 3,957 | 73,557 | 2,413 | – | – |
| | [300, 500] | 45,139 | 44,322 | – | 79,778 | 79,442 | – | 49,539 | – | – |
| | [100, 500] | 23,530 | 22,864 | – | 42,563 | 41,731 | – | 26,722 | – | – |
| 100 | [30, 120] | 1,018 | 969 | 237,194 | 8,891 | 10,637 | 110,096 | 1,717 | – | – |
| | [300, 500] | 24,053 | 23,647 | – | – | 56,324 | – | 23,477 | – | – |
| | [100, 500] | 8,766 | 8,567 | – | 22,646 | 22,403 | – | 10,886 | – | – |
| *Device failure rate 1%* | | | | | | | | | | |
| 6 | [30, 120] | 34,051 | 215,208 | 52,632 | 193,609 | 244,350 | 47,612 | 15,951 | 64,861 | – |
| | [300, 500] | – | – | – | – | – | – | – | – | – |
| | [100, 500] | 268,187 | – | – | – | – | – | 181,425 | – | – |
| 30 | [30, 120] | 69,286 | – | 144,148 | – | – | 94,139 | 9,496 | 280,270 | – |
| | [300, 500] | 262,768 | – | – | – | – | – | 115,880 | – | – |
| | [100, 500] | – | – | – | – | – | – | 65,915 | – | – |
| 100 | [30, 120] | 173,068 | – | – | – | – | 139,263 | 10,031 | – | – |
| | [300, 500] | 291,705 | – | – | – | – | – | 74,214 | – | – |
| | [100, 500] | – | – | – | – | – | – | 40,334 | – | – |

performance with line-based aggregation.

**Tree-based aggregation.** Without device failures, both Pulsar and FreeBeacon can finish data aggregation under all considered scenarios, while Find is only able to finish three out of nine. The completion time by FreeBeacon is mostly better than that by Find. For example, in the scenario with charging time range [30, 120] slots and six devices, FreeBeacon outperforms Find by almost 5.9×. However, under 1% device failures, Find performs the best among all approaches where it manages to complete three scenarios while both Pulsar and FreeBeacon can only finish one. For the scenario completed by both Pulsar and FreeBeacon, FreeBeacon still has superior performance.

**Ring-based aggregation.** Under no device failures, FreeBeacon outperforms the baselines to a large extent, where it finishes all scenarios while Pulsar finishes only one and Find does not finish any of the scenarios. For the scenario finishes by both Pulsar and FreeBeacon, we can see that FreeBeacon outperforms Pulsar by a factor as large as 7.88. Under 1% device failures, Find does not finish any scenario, Pulsar finishes only one, while FreeBeacon finishes eight out of the nine scenarios. The significant performance gain of FreeBeacon can be attributed to its excellent ability to handle device failures and collisions, which is extremely important for the ring-based aggregation schedule since many concurrent transmissions happen in every aggregation round.

### B. Beacon Discovery Performance

We now focus on the performance of the beacon discovery protocol of FreeBeacon. Here, we choose Pulsar as our main baseline since Pulsar also uses a battery-powered device that plays a similar role as the beacon in FreeBeacon. However, the discovery in Pulsar is done with a simple linear probing approach. The other baselines do not have the discovery process, and hence, they are not included in this comparison. We set $T_{dist} = 30$ slots and $T_b = 7$ slots for FreeBeacon and for Pulsar we use a cycle of 31 slots for the coordinator; Pulsar allocates a dedicated slot for the coordinator. Experiments are done with Python-based numerical simulations and the numbers are gathered from 50 independent runs.

Figure 10 depicts the discovery time for Pulsar and FreeBeacon, under varying energy conditions and varying numbers of devices (2, 12, and 30). The discovery time is measured as the time it takes for all battery-free devices to finish discovering the battery-powered device. When the number of battery-free devices is small, Pulsar has better performance than FreeBeacon. This is expected since linear probing is quite effective when the collision rate is low, which is the case with a small number of battery-free devices. However, when the number of devices increases, we can see that FreeBeacon starts to outperform Pulsar. This is due to the meticulous design of the number-theory-based mechanism, providing discovery guarantees while avoiding collisions. The results demonstrate that the beacon discovery protocol in FreeBeacon is effective, especially for large-scale scenarios.

### C. Pairwise Communication Performance

To understand the performance of FreeBeacon under different communication tasks, we consider a pairwise communication pattern where we schedule devices into pre-defined pairs and ask devices to find their counterpart.

*1) Experiments on a Controlled Testbed:* To further evaluate FreeBeacon with more realistic setups, we build a controlled testbed with two sets of FreeBeacon prototypes using hardware to emulate the behavior of battery-free devices. Each prototype consists of the following components: two TI-MSP430FR5994 microcontrollers serving as the main control and power supply, respectively, the intermittency is emulated by replaying existing charging time traces through a MOS-FET controlled by the microcontroller. We use a NORDIC-NRF52840 board for communication with BLE. The beacon is emulated with a NORDIC-NRF52840 board. An overview of the setup can be seen in Figure 9(right).

We consider two scenarios with 0% and 10% device failure rates and use the office energy traces. According to the setup, we set the length of a slot to 410 ms considering the time needed for initialization and communication. We set the size of the distribution cycle to $T_{dist}$ to 30 slots and the beacon cycle $T_b$ to 7 slots. Figure 11 shows the results where we can observe that FreeBeacon outperforms Find significantly. In the case of no device failures, the performance gain is around

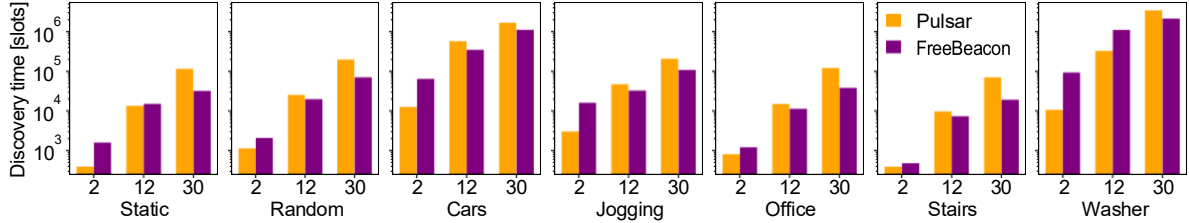

Fig. 10: Overall discovery performance of FreeBeacon compared with baseline Pulsar.

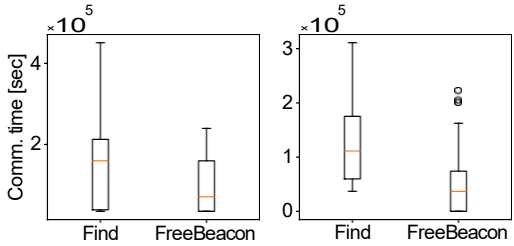

Fig. 11: Pairwise comm. performance on controlled testbed (left) w/o and (right) w/ 10% device failures.

33.3% while it increases to 59.5% in the case of 10% device failure rate. The results are in line with those from simulations and further prove the superior capability of managing device failures and handling the resulting collisions.

*2) Numerical Simulations:* We set $T_{dist}$ = 30 slots and $T_b$ = 7 slots for FreeBeacon. Experiments are done with Python-based numerical simulations and the numbers are gathered from 50 independent runs. Figure 12 shows the overall performance of FreeBeacon under the condition that no device failures happen. We vary the number of participating devices (2, 12, and 30) and compare the communication time which is the time for each approach to complete all device pairing. Overall, we can observe a clear performance gain from hybrid approaches (both Pulsar and FreeBeacon) when compared with random-guess-based approaches. For example, in the scenario of Cars, FreeBeacon achieves a tremendous communication time reduction of up to 86.1× and 8.75× when compared with Find and Flync-Find, respectively. This performance gain can be consistently observed under all environmental conditions and with different numbers of devices. When we compare FreeBeacon with Pulsar, we notice that the performance of the two approaches is comparable. But in more chaotic conditions (e.g., Cars) where the trace shows a higher variance in the charging time, thanks to its careful design in avoiding conflicts, FreeBeacon still outperforms Pulsar, although the gain is limited. We will show a much larger gain when device failures occur.

Figure 13 shows that FreeBeacon achieves significantly better performance when device failures occur. Here, we set the number of devices to 30 and the most promising distribution cycle size $T_{dist}$ = 51 as shown from the previous experiment. We vary the device failure rate among 1%, 5%, and 10% and repeat the same experiment as above. The results show that FreeBeacon outperforms all baselines with impressive gains in communication time reduction, e.g., 264.37×, 26.27×, and 4.82× when compared with Find, Flync-Find, and Pulsar, respectively. This performance improvement can be mostly attributed to FreeBeacon's ability to handle device failures and avoid rediscovery overheads.

### D. Impact of Parameters

*1) Impact of the Beacon Cycle.:* The selection of the beacon cycle highly depends on the scenario, but we notice that a smaller beacon cycle generally leads to better communication performance. To verify this observation, we choose different distribution cycles (i.e., 30, 51, 100, 120, 160 slots) and set the beacon cycle to the coprimes of the distribution cycle in the range of [2, 200]. The charging time is drew from [1,500] slots as this is the typical range. We evaluate the synchronization time between two battery-free devices. Figure 14 shows that under all distribution cycles, the synchronization time shows a linearly increasing trend, confirming our observation. In practice, we should always select a beacon cycle size that is the smallest coprime to the distribution cycle.

*2) Impact of the Distribution Cycle.:* The selection of the distribution cycle has a large impact on the performance of FreeBeacon. To understand this impact, we first consider three different scenarios where we sweep through the distribution cycle space. For the three scenarios, we set the device charging time as follows: fixed to 300 slots statically, randomly drawn from the range [1, 500] slots, and randomly drawn from the range [300, 500] slots. We consider the communication between two battery-free devices in this experiment. We vary the distribution cycle from 2 to the maximum charging time. Figure 15 shows the results. In the static case, the communication time first reduces (almost) linear with the increase of the distribution cycle. However, when the distribution cycle reaches the maximum charging time, the performance increases significantly. In other cases, we notice that the performance is unstable when the distribution cycle is smaller than the smallest charging time, but increases when it approaches the charging time range.

The above results show that finding the best parameter for FreeBeacon is not easy for different scenarios. Instead of (over-)optimizing this parameter, we decide to choose the best parameter from three manually picked values: 30, 51, and 100. Figure 16 shows the results of FreeBeacon under these distribution cycle sizes. As we can generally observe $T_{dist}$ = 51 provides the best average performance under all the environmental conditions in general. For example, under the Office scenario, $T_{dist}$ = 51 outperforms $T_{dist}$ = 30 and $T_{dist}$ = 100 by 25.89% and 26.82% on average, respectively.

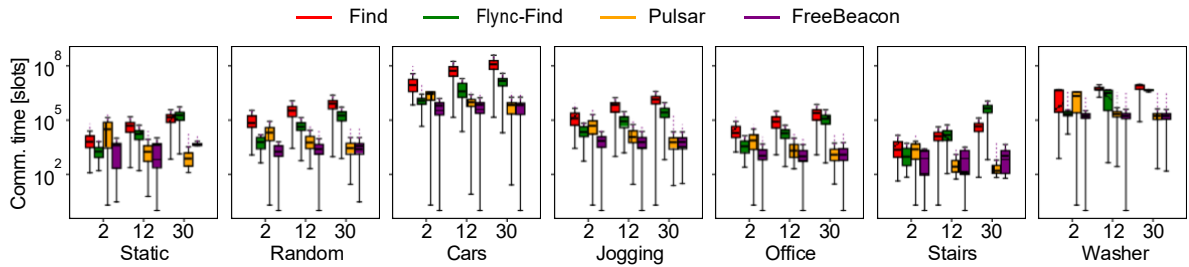

Fig. 12: Pairwise communication performance of FreeBeacon without device failures.

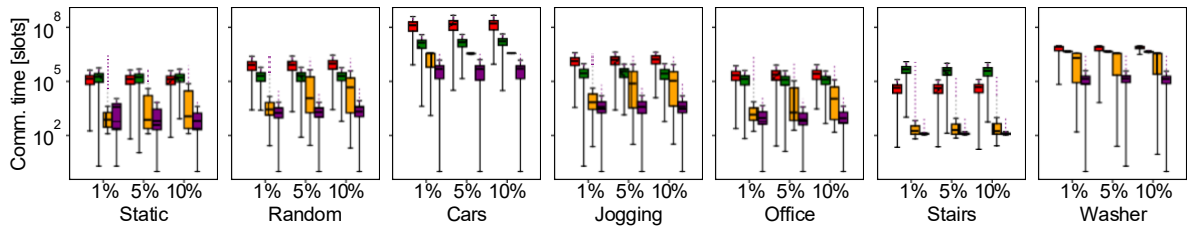

Fig. 13: Pairwise communication performance of FreeBeacon under varying device failure rates.

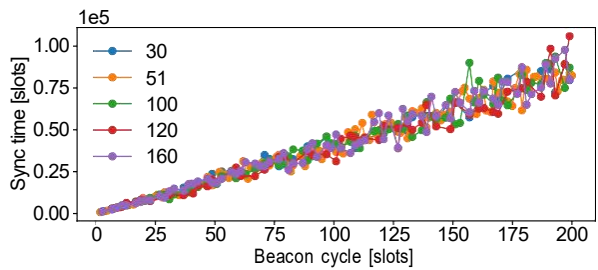

Fig. 14: Impact of the beacon cycle size $T_b$ under different distribution cycles.

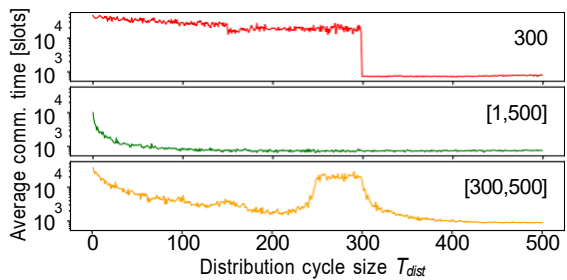

Fig. 15: Impact of the distribution cycle size $T_{dist}$ under specific charging time range.

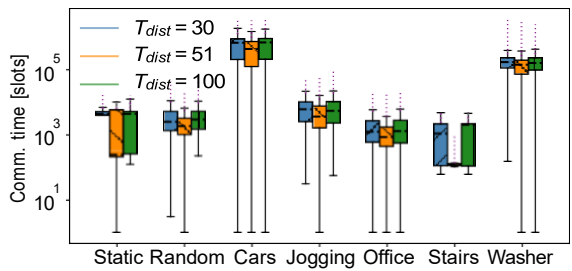

Fig. 16: Impact of the distribution cycle size $T_{dist}$ on real-world charging time traces.

Hence, we will use $T_{dist} = 51$ as the default distribution cycle configuration for FreeBeacon.

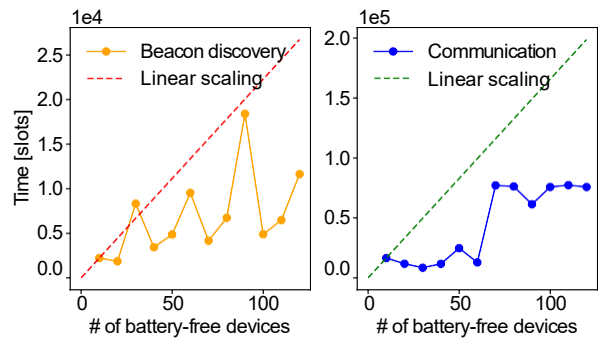

Fig. 17: Beacon discovery (left) and communication (right) performance under varying numbers of devices.

*3) Scalability:* We also evaluate the scalability of Free-Beacon by gradually increasing the number of battery-free devices and measuring the beacon discovery and the pairwise communication time. The charging time of all devices is sampled from the $[1, 500]$ slots range. Figure 17 shows that both the beacon discovery and device communication time stay mostly below the linear scaling line, demonstrating good scalability of FreeBeacon.

## V. LIMITATIONS AND DISCUSSION

In this section, we discuss the limitations of FreeBeacon and point out some directions for future work.

**Supporting multiple beacons.** Currently, FreeBeacon only supports one beacon by which all battery-free devices are coordinated. While we can always increase the distribution cycle to accommodate more devices, the coordination overhead will increase dramatically and the system's efficiency will inevitably drop when there are too many devices. A straightforward solution would be to arrange battery-free devices into clusters and assign a separate beacon for each cluster. A more practical approach is to let battery-free devices discover and join the coordination of the beacon that it sees first. In either case, communication across devices coordinated by different beacons is still an open challenge.

**Distribution cycle adaptation.** The distribution cycle is decided before deployment currently and its adaptation is not supported. Yet, there are cases where the distribution cycle size needs to be adjusted on the fly: (1) When devices join and leave the system dynamically, the distribution cycle should be tuned based on that. (2) The environmental condition changes dramatically, calling for adjusting the distribution cycle to optimize system efficiency. Supporting such adaptation is not easy, as the beacon requires significantly more involvement. We leave it for future work.

**Large-scale experiments with field tests.** While our evaluation involves several methods covering both testbed-based experiments and simulations, our testbed is restricted to a small scale and cannot be used to test the scalability of the proposed algorithms due to cost limitations. We will keep looking for opportunities to extend our testbed and perform more large-scale experiments on it.

## VI. Related Works

We discuss related works in the area of sustainable IoT.

**Energy-harvesting WSNs.** To sustain traditional WSNs, researchers have explored the concept of energy harvesting for charging the battery or super-capacitor on wireless sensors to prolong their life. Based on this concept, different energy management, topology construction, and communication scheduling mechanisms have been proposed [3], [27], [28], [29]. However, energy-harvesting wireless sensors have fundamental differences from battery-free devices because the former has predictable power supplies while the latter may suffer from random power failures depending on the uncertain energy availability in the environment. Due to such core differences, existing approaches for energy-harvesting WSNs are not directly applicable to battery-free IoT.

**Intermittent Computing.** The intermittency of battery-free devices calls for software support for programs running on such devices to make forward progress. The challenge comes from the fact that a battery-free device, when waking up, may only be able to sustain its operation very briefly (e.g., a few milliseconds) before falling into the charging state again. A program execution is unlikely to finish within this tiny amount of time, hence the program execution state needs to be preserved, e.g., on non-volatile memories such as FRAM before every power failure, so the state can be recovered and execution continues when the device wakes up again. There is a long list of works in this line [30], [31], [32], [33], [34], [8], [7], [35], [36], [37], [18], [38], [39], [40], [41], [42], [43], [44], [13], [45], [16], [46], [47], [48], [49], [50], [51], [52]. Generally, there are two categories: Checkpoint-based approaches, which insert checkpoints into the program, allowing the runtime system to save program states at these checkpoints during execution [15], [53], [54]. (2) Task-based approaches, which transform the program into tasks that can be re-executed after a power failure without consistency issues [55], [14], [17]. All these works, while being orthogonal, provide a solid foundation for our work.

**Communication for battery-free IoT.** There are generally two approaches for battery-free communication: backscatter [56], [57], [58], [7] and radio-based [59], [60]. Related to ours are the radio-based approaches where people have explored synchronization and communication between battery-free devices [5], [12], [11], [24], [20], [25]. Most of these works rely on random guesses or linear probing, which are not efficient in challenging environments and suffer from robustness issues. The closest to ours is Pulsar, which is also based on a hybrid design. While the high-level idea is similar, the core design is completely different (e.g., Pulsar is based on linear probing while FreeBeacon is based on a more principled approach inspired by number theory). More importantly, Pulsar does not deal with data aggregation as we do in this paper.

## VII. Conclusions

In this work, we presented FreeBeacon, a novel scheme designed to tackle the communication challenges inherent to battery-free IoT devices, driven by the unpredictable nature of ambient energy harvesting. By introducing a small number of battery-powered beacons to provide limited certainty, FreeBeacon simplifies the complexities associated with intermittent device communication. Our approach leverages this certainty to enable battery-free devices to interact robustly with the beacon and follow efficient communication schedules, leading to more effective data aggregation. Extensive simulation and prototype-based experiments demonstrate that FreeBeacon outperforms state-of-the-art methods, reducing data aggregation times by an order of magnitude. These results suggest that FreeBeacon offers a promising direction for enhancing the sustainability of future IoT systems reliant on battery-free devices, making them more attractive for real-world deployment.

## VIII. APPENDIX

### A. *FreeBeacon* Protocols

---

**Algorithm 1** Beacon discovery protocol

---

1: **procedure** BEACON($T_{dist}$, $T_b$)  ▷ Beacon logic
2:   $n \leftarrow 1$
3:   **while** true **do**  ▷ Loop once upon every wake-up
4:     $t_b \leftarrow (n \times T_b)\%T_{dist}$
5:     Broadcast $t_b$ upon receiving a request
6:     $n \leftarrow n + 1$
7: **procedure** DEVICE($T_{dist}$, $T_i$, $t_i$)  ▷ Device logic
8:   $count \leftarrow 0$
9:   **while** true **do**  ▷ Loop once upon every wake-up
10:     $count \leftarrow count + 1$
11:     $delay \leftarrow T_{dist} - T_i\%T_{dist}$
12:     **if** $count > T_{dist}$ **then**  ▷ Handle conflicts
13:       $delay \leftarrow delay + T_{dist} \times$ RANDOM$(0, 1)$
14:       Delay by $delay$ slots
15:     $t_b \leftarrow$ SENDRECEIVE()  ▷ Discovery attempt
16:     **if** $t_b \neq 0$ **then**  ▷ Successful discovery
17:       **if** $t_b! = t_i$ **then**
18:         $b_i \leftarrow (t_i - t_b)\%T_{dist}$  ▷ Slot correction
19:       $delay \leftarrow (T_{dist} - T_i\%T_{dist} + b_i)\%T_{dist}$
20:       Delay by $delay$ slots from now on
21:       Break

---

**Algorithm 2** Device-to-device communication protocol

---

1: **procedure** BEACON($T_{dist}$, $T_b$)  ▷ Beacon logic
2:   $n \leftarrow 1$
3:   **while** true **do**  ▷ Loop once upon every wake-up
4:     $t_b \leftarrow (n \times T_b)\%T_{dist}$
5:     Broadcast $t_b$ upon receiving a message where $t_{recv} \neq t_b$
6:     $n \leftarrow n + 1$
7: **procedure** SENDER($T_{dist}$, $b_{snd}$, $T_{snd}$, $t_{snd}$, $t_{recv}$)  ▷ Sender logic
8:   $delay \leftarrow (T_{dist} - T_{snd}\%T_{dist} + b_{snd})\%T_{dist}$
9:   $distance \leftarrow (t_{recv} - t_{snd})\%T_{dist}$
10:   $delay \leftarrow delay + distance$
11:   **while** ture **do**  ▷ Loop once upon every wake-up
12:     Delay by $delay$ slots
13:     $response \leftarrow$ SENDRECEIVE(data)
14:     **if** $response.src$ is $receiver$ **then**  ▷ Success
15:       Delay by $(T_{dist} - delay)\%T_{dist}$  ▷ Rollback
16:       Break
17:     **else if** $response.src$ is $beacon$ **then**
18:       $b_{snd} \leftarrow (t_{snd} - t_b)\%T_{dist}$  ▷ Slot correction
19: **procedure** RECEIVER($T_{dist}$, $b_{snd}$, $T_{snd}$)  ▷ Receiver logic
20:   $delay \leftarrow (T_{dist} - T_{snd}\%T_{dist} + b_{snd})\%T_{dist}$
21:   **while** true **do**  ▷ Loop once upon every wake-up
22:     Delay by $delay$ slots
23:     $data \leftarrow$ RECEIVESEND(ack)

---

### B. Proof to Theorem 1

*Proof.* The proof can be conducted by showing that the index $t_b$ forms a Weyl sequence [22] when $T_b$ and $T_{dist}$ are co-prime. Following the equidistribution theorem, a Weyl sequence satisfies that given an integer $k$ relatively prime to an integer modulus $m$, the sequence of all multiples of $k$ (i.e., $0, k, 2k, ...$) is equidistributed modulo $m$. That means that the elements in the sequence will be uniformly distributed in the interval $[0, m)$. The beacon index naturally forms a Weyl sequence if we set $k = T_b$ and $m = T_{dist}$, where $T_b$ and $T_{dist}$ are co-prime by design. Hence, the beacon index $t_b$ will visit each slot of the distribution cycle with the same frequency. In other words, no matter when the device wakes up, there exists at least one slot where the device and the beacon wake up at the same time, which concludes the proof. □